\documentclass[12pt]{iopart}
\usepackage{graphicx}
\usepackage{bm}

\newcommand{\eqb}{\begin{eqnarray}}
\newcommand{\eqe}{\end{eqnarray}}
\newcommand{\diff}{{\rm d}}
\newcommand{\ppar}{{p_\|}}

\newcommand{\pperp}{{p_\bot}}

\newcommand{\mnras}{MNRAS}
\newcommand{\aap}{Astronomy \& Astrophysics}
\newcommand{\apj}{Astrophysical J.}

\begin{document}

\title{Waves in pulsar winds}

\author{J.G. Kirk}

\address{Max-Planck-Institut f\"ur Kernphysik,
Postfach 103980, 69029 Heidelberg, Germany} 
\ead{john.kirk@mpi-hd.mpg.de}
\begin{abstract}
  The radio, optical, X-ray and gamma-ray nebul\ae\ that surround many
  pulsars are thought to arise from synchrotron and inverse Compton
  emission. The energy powering this emission, as well as the magnetic
  fields and relativistic particles, are supplied by a \lq\lq
  wind\rq\rq\ driven by the central object. The inner parts of the
  wind can be described using the equations of MHD, but these break
  down in the outer parts, when the density of charge carriers drops
  below a critical value. This paper reviews 
the wave properties of the inner part
  (striped wind), and uses  a relativistic two-fluid model
  (cold electrons and positrons) to re-examine the nonlinear
  electromagnetic modes that propagate in the outer parts.  It is
  shown that in a radial wind, two solutions exist for circularly
  polarised electromagnetic modes. At large distances one of them
  turns into a freely expanding flow containing a vacuum wave, whereas
  the other decelerates, corresponding to a confined flow.
\end{abstract}

\maketitle

\section{Introduction}
The size and quality of the observational database relating to pulsar
wind nebul\ae\ has increased dramatically in the past five years,
stimulating research into both the modelling of their spectra and
the theory of the winds themselves (for recent reviews see
\cite{gaenslerslane06,kirketal09}). The basic idea of
the pulsar wind as a mix of particles, fields and waves, 
was outlined over thirty years ago \cite{reesgunn74}. Close to the pulsar,
the wind moves outwards at relativistic speed. 
When the the ram-pressure drops 
to approximately that of the surroundings, a \lq\lq termination shock\rq\rq\ 
occurs, and strong synchrotron and inverse Compton radiation is emitted
in the relatively slowly moving plasma in the surrounding \lq\lq nebula\rq\rq. 
However, although 
progress has been made on understanding the structure of the nebula,
\cite{komissarovlyubarsky04,delzannaetal04,delzannaetal06}
there are still difficulties associated with the propagation of the wind.
For example, the questions of where and how the Poynting flux that it 
carries is dissipated into particles remains open 
\cite{melatosmelrose96,melatos98,kirkskjaeraasen03,petrilyubarsky07,lyutikov10}.
This paper, does not present a solution to these problems, but
concerns itself with an analysis of two simple, nonlinear waves that
propagate in the wind, and are likely to play a crucial role in
shaping the global solution. 

A rotating magnetised neutron
star can be expected to emit predominantly linearly polarised waves in the 
equatorial plane, and circularly polarised waves along the rotation axis.
In addition, the waves in the equatorial region may carry a non-vanishing
phase-averaged component of the magnetic field component. At intermediate 
latitudes, the waves will, in general, be elliptically polarised and may 
also carry a non-oscillatory component. A linear analysis of these waves
is clearly inadequate, since the characteristic quiver frequency of an 
electron at points interior to the termination shock front is
many orders of magnitude larger than the rotation frequency of the star. 
A nonlinear analysis, however, is feasible only for a few simple modes. 

The first mode we select for analysis is  
a linearly polarised wave --- the {\em striped wind} --- 
which has vanishing phase-averaged magnetic field.
This wave can
be thought of as an entropy wave or, alternatively, a rotational
discontinuity. It requires a minimum plasma density in
order to propagate, and so can exist only up to some maximum distance
from the pulsar. Its subluminal phase velocity equals its group
velocity, and the field is essentially magnetostatic in the wave
frame. Since it propagates at highly super-magnetosonic velocity, 
its evolution is fixed by boundary conditions
close to the star. This wave is likely to provide a good description 
of the interior zone of a pulsar wind in the equatorial regions.

Following this, the propagation of nonlinear electromagnetic
waves of superluminal phase velocity is addressed. These waves cut
off if the plasma density exceeds a certain limit, and so propagate
only beyond some minimum distance from the star. Both 
linearly and circularly polarised nonlinear plane waves can be constructed
analytically, but the formalism is much simpler for the case of 
circular polarisation. In this paper, a new analysis is
presented of circularly polarised 
waves in spherical geometry, using the short wavelength
approximation. This is appropriate in the polar regions
of a pulsar wind. However, one can expect that the linearly polarised 
electromagnetic modes in the equatorial region exhibit qualitatively
similar behaviour. Because the phase speeds are superluminal, 
both inner and outer boundary conditions are required to
specify the wave structure. At large radius, two types of solution are found: 
in one of them the flow velocity
stagnates in a region of constant density, whereas, in the other, it
asymptotes to constant radial velocity. The former applies to 
pulsar wind nebulae, which are confined by an external
medium. The latter applies to unconfined winds.

\section{Parameters of the wave}

Consider radially propagating waves in a region far 
outside the light-cylinder of the pulsar, 
i.e., $\hat{r}=r/ r_{\rm L}\gg1$, 
where $r_{\rm L}=cP/2\pi$ is the radius of the light cylinder 
and $P$ is the pulsar period. In this region, 
the magnetic field is approximately transverse, 
because the longitudinal (radial) component 
decays as $\hat{r}^{-2}$, 
whereas transverse component decays as $\hat{r}^{-1}$. Furthermore,
a plane wave approximation is appropriate, since 
the wavelength is small compared to the curvature of the wavefront. 
Though approximately radial, pulsar winds are definitely not 
spherically symmetric, which is accounted for by specifying a finite
solid angle $\Omega_{\rm s}<4\pi$ in which the wind propagates. 

The wave properties depend on 
the phase-averaged radial particle flux density $\left<J\right>$,
the phase-averaged total
radial energy flux density $\left<F\right>$, 
(which is the component
$T^{0,1}$ of the sum of the stress-energy tensors of the 
fields and particles), and the 
phase-averaged flux density of radial momentum, $\left<Q\right>$,
(which is  
the component $T^{1,1}$). This latter quantity depends on both 
$\left<F\right>$ and on the way in which the energy flux is shared 
between the particles
and fields. For example, for a vacuum wave, the momentum flux density
equals the energy flux density/c, whereas for cold particles 
of speed $v$ in the absence of fields the momentum density equals $v/c^2$ times 
the energy flux density. 

Instead of the fluxes, however, it is more
convenient to use dimensionless parameters.
Conventionally, a steady, relativistic MHD wind is 
characterised by the parameter $\mu$ introduced by 
Michel \cite{michel69},
which is the luminosity carried
by the wind per unit rest mass. It gives the 
Lorentz factor each particle would have if all the energy flux were 
carried  by the particles:
\eqb
\mu&=&\frac{\left<F\right>}{mc^2 \left<J\right>}
\label{defmu}
\eqe
In a pulsar wind, $\mu$ is constant, (provided pairs are not created 
or annihilated and no energy is dissipated into radiation 
in the region under consideration), but its
value is uncertain. For the Crab Nebula the average of $\mu$ 
over the pulsar lifetime
can be found by combining an estimate of the number of
radiating electrons/positrons, obtained from the synchrotron luminosity, 
with an estimate of the average magnetic field, obtained from the 
gamma-ray luminosity. But the current value 
is thought by many to be rather higher than in the past. 
One estimate is $\mu\sim10^4$ \cite{kirketal09} but the real value 
might be larger.

The distribution of energy flux between particles and fields 
can be described by the magnetisation parameter $\sigma$, 
defined, in an MHD flow, as the ratio of the enthalpy density of the magnetic
field to that of the plasma, measured in the plasma rest frame.
This quantity depends on the fraction of the pulsar luminosity that goes into
creating pairs, i.e., not only on their number, but also on the Lorentz factor
with which they are injected into the wind.
$\sigma$ is the ratio of the maximum
possible Lorentz factor (the parameter $\mu$) to this injection Lorentz
factor. This is generally thought to be 
high $\sim10^2$--$10^3$, so 
that $\sigma\approx 10^{-2}\mu$--$10^{-3}\mu$.

Finally, as a dimensionless expression of the 
total energy flux it is convenient to choose the 
strength parameter 
of a circularly 
polarised vacuum wave that carries the same energy flux as the 
nonlinear wave.
Writing 
$E_{\rm eff}$ for the magnitude of the 
electric field in this wave, the strength parameter is defined as
$a={eE_{\rm eff}}/{(mc\omega)}$,
and the required value of $E_{\rm eff}$ is determined from
$c{E_{\rm eff}^2}/{(4\pi)}=\left<F\right>$:
\eqb
a&=&\left[\frac{4\pi e^2 \left<F\right>}{m^2c^3\omega^2}\right]^{1/2}
\label{defa}
\eqe
In a pulsar, this quantity varies with position in the wind:
$a=a_{\rm L}/\hat{r}$
where $a_{\rm L}$ is the 
strength parameter evaluated at $r=r_{\rm L}$:
\eqb
a_{\rm L}&=&\left[\frac{4\pi e^2 L}{m^2c^5\Omega_{\rm s}}
\right]^{1/2}
\nonumber\\
&=&4.3\times 10^{10} L_{38}^{1/2}
\left(4\pi/\Omega_{\rm s}\right)^{1/2}
\label{aldefinition}
\eqe
where $L$ is the luminosity carried by the (radial) wind,
and $L_{38}=L/(10^{38}\,\textrm{erg\,s}^{-1})$.
For the Crab, $L=4.6\times10^{38}\,\textrm{erg\,s}^{-1}$; 
many pulsars have $L=10^{31}$ --- $10^{33}\,\textrm{erg\,s}^{-1}$, 
and some go down to 
$10^{28}\,\textrm{erg\,s}^{-1}$. 
The value of $\Omega_{\rm s}$ is not known, but is often 
assumed to be $\sim1$.
In a steady, radial wind, 
\eqb
\hat{r}^2\left<J\right>&=&C_J
\\
\hat{r}^2\left<F\right>&=&mc^2 C_F
\\
\hat{r}^2\left<Q\right>&=&mc C_Q
\eqe
where $C_{J,F,Q}$ are constants
and $\mu=C_F/C_J$. The particle density is frequently expressed in units
of the \lq\lq Goldreich-Julian\rq\rq\ density at the light cylinder
\cite{lyubarskykirk01}. For a magnetically dominated, relativistic flow,
the corresponding 
\lq\lq multiplicity\rq\rq\ parameter is $\kappa=a_{\rm L}/(4\mu)$.

\section{The striped wind --- an MHD wave}

Coroniti \cite{coroniti90} and Michel \cite{michel94} suggested
idealised solutions in which the wind settles down into stripes of
magnetic flux of alternating polarity containing cold plasma,
separated by current sheets containing hot plasma.  
In this solution, the plasma is at rest in a reference frame that,
with respect to the star, moves radially outwards with Lorentz factor
$\gamma_{\rm MHD}=\left(1-\beta_{\rm MHD}^2\right)^{-1/2}$. 
In the comoving frame, the electric field vanishes, and
the magnetic field is toroidal. There are two current 
sheets per wavelength.

Such solutions
can exist only if the wind carries a sufficient number of particles to
carry the current that Ampere's law requires must flow in the
sheets. To within a factor of the order of unity, this condition is
equivalent to the requirement that particles can be confined in the
sheet, i.e., that the gyro-radius of a hot particle moving in the
magnetic field present in the cold part of the flow is smaller than the sheet
thickness. It was noticed in \cite{coroniti90} that this condition
implies a minimum dissipation rate: as the 
density decreases radially, the minimum sheet thickness increases.
According to \cite{lyubarskykirk01}, dissipation
forces the wind to accelerate, which, because of time-dilation,
reduces the apparent dissipation rate observed in the lab.\ frame.
Of course, the actual dissipation rate is unknown. If it lies below the 
minimum, then the solution moves out of the range of validity of the 
MHD equations. If it lies above the minimum, the sheet 
expands, consuming 
the surrounding cold plasma and annihilating the field embedded in it.
In the equatorial plane, adjacent sheets carry equal and opposite
magnetic fluxes, and it is conceivable that the wave eventually
annihilates the entire magnetic field. However, at higher latitudes,
a residual flux remains after removal of 
the alternating component. 

Several possible dissipation or \lq\lq reconnection\rq\rq\ 
scenarios were investigated 
in \cite{kirkskjaeraasen03}. The conclusion, 
in the case of the Crab, is that the maximum possible
dissipation rate, combined with a very high particle content
(small $\mu$), could
marginally enable the conversion of all the wind energy flux into
particle kinetic energy before the termination shock is encountered.
Given a reconnection prescription, 
self-similar solutions can be found that describe how the flow accelerates.
For example, for dissipation at the minimum rate, the radial dependence can be 
approximated by \cite{kirkskjaeraasen03}
\eqb
\gamma_{\rm MHD}&=&
\mu \sqrt{3\hat{r}/(\pi a_{\rm L})}
\label{gammamhd}
\eqe
This relation holds in the supersonic part of the flow 
($\gamma_{\rm MHD}>\mu^{1/3}$), 
and before the magnetic field is 
annihilated completely ($\gamma_{\rm MHD}<\mu$). 
In terms of the radius, these restrictions
imply $a_{\rm L}>\hat{r}>a_{\rm L}/\mu^{4/3}$. 

However, there exists an additional, independent
constraint that is imposed by the use of a short-wavelength 
approximation in this analysis. This requires that the 
striped wind pass
through a series of quasi-stationary states as it expands in the 
radial geometry. For this to remain valid, the gyro-period of 
particles contained in the sheets must be short compared to the 
expansion timescale. An equivalent formulation of this condition 
is that the speed at which the sheet expands must remain subsonic.
This implies an upper limit
on the wind Lorentz factor, and, hence, the radius:
\eqb
\gamma_{\rm MHD}&<&\gamma_{\rm M}\,\approx\,a_{\rm L}/\mu
\qquad
\hat{r}\,<\,a_{\rm L}^3/\mu^4
\eqe
above which the wave can no longer be described by the MHD equations.

In order to relate the striped wind solutions, which apply in the inner 
part of the wind, to the electromagnetic waves to be described in the
following section, which apply in the outer part of the wind, it
is useful to write down the phase-averaged
fluxes carried by the wind, assuming the 
contribution of the current sheets to be negligible.
Noting that the electric and magnetic fields are related by
$E=\beta_{\rm MHD}B$, are orthogonal, and are constant in between
the current sheets, (although they change sign across each sheet), and 
that the wave speed equals the fluid 
speed, i.e., $\gamma_{\rm
  MHD}=\gamma=\sqrt{1+\ppar^2}$, where $\gamma$ is the particle
Lorentz factor and $\ppar$ the radial momentum in units of $mc$,
one finds: 
\eqb \left<J\right>&=&2n\ppar c
\\
\left<F\right>&=&2nmc^3\ppar\gamma + c\beta_{\rm
  MHD}\frac{B^2}{4\pi}
\\
\left<Q\right>&=&2nmc^2\ppar^2 + \left(1+\beta_{\rm
    MHD}^2\right)\frac{B^2}{8\pi} 
\eqe 
where $n$ is the proper number
density of electrons, which equals that of positrons.  
In terms of the magnetisation parameter, $\sigma=B^2/\left(8\pi\gamma_{\rm MHD}^2nmc^2\right)$: 
\eqb
\frac{\left<F\right>}{mc^2\left<J\right>}
&=&\frac{C_F}{C_J}
\,\equiv\,\mu\,=\,\gamma_{\rm MHD}\left(1+\sigma\right)
\label{mhdcfcj}
\\
\frac{\left<Q\right>}{mc\left<J\right>}
&=&\frac{C_Q}{C_J}
\,=\,\left(\gamma_{\rm MHD}^2-1\right)^{-1/2}
\left[\gamma_{\rm MHD}^2\left(1+\sigma\right)
-\left(1+\frac{\sigma}{2}\right)\right]
\label{mhdcqcj}
\eqe
and so in this case, in which the effect of the current sheets 
is neglected, both $\gamma_{\rm MHD}$ and $\sigma$ remain constant in the flow. 
For $\mu\gg1$ and $\sigma\ll\mu^2$, the energy and momentum flux densities
are almost equal:
\eqb
\frac{C_F}{C_J}&=&\mu;\qquad
\frac{C_Q}{C_J}\,\approx\,\mu-\frac{1+\sigma}{2\mu}
\label{cqcjmhd}
\eqe

\section{Electromagnetic waves}

A relatively simple description of the outer parts of 
a pulsar wind can be based on a
model in which two cold, relativistic, 
fluids (electrons and positrons) propagate
radially outwards in transverse electromagnetic fields. 
Strong waves in systems similar to this were analysed by 
\cite{maxperkins71,maxperkins72,kennelschmidtwilcox73,clemmow74}, using the 
method of \cite{akhiezerpolovin56}. In connection with 
pulsars \cite{asseokennelpellat75} and \cite{asseopellatllobet84} 
investigated strong, linearly polarised waves in spherical geometry. 
Here the effect of spherical geometry on
circularly polarised electromagnetic waves is analysed.
 
Under the above assumptions, the proper number 
density is the same for each fluid:
$n_+=n_-=n$. In spherical polar coordinates $(r,\theta,\phi)$ 
their radial velocity is also the same: 
$v_{r+}=v_{r-}=v_r$
and the toroidal and azimuthal velocities are equal in magnitude but of 
opposite sign: $v_{\theta+}=-v_{\theta-}$, $v_{\phi+}=-v_{\phi-}$.  
It proves convenient to use the Lorentz factor of the fluids
$\gamma_+=\gamma_-=\gamma=c/\sqrt{c^2-v_{r}^2-v_{\theta+}^2-v_{\phi+}^2}$,
to define the dimensionless radial momentum:
$\ppar=v_{r}\gamma/c$, and to introduce complex quantities 
representing the dimensionless transverse momentum:
$\pperp=\left(v_{\theta+}+iv_{\phi+}\right)\gamma/c$
($=-\left(v_{\theta-}+iv_{\phi-}\right)\gamma/c$) and the 
electric and magnetic fields:
$E=E_{\theta}+iE_{\phi}$, $B=B_{\theta}+iB_{\phi}$.
Then the equations to be solved are the 
continuity equation
\eqb
\frac{\partial}{\partial t}\left(n\gamma\right)+
\frac{c}{r^2}\frac{\partial}{\partial r}\left(r^2 n\ppar\right)&=&0
\label{circcontinuity}
\eqe
the fluid equations of motion 
\eqb
\frac{\diff \ppar}{\diff\tau}&=&-\frac{e}{mc}
\textrm{Im}\left(\pperp B^*\right)
\label{circu}\\
\frac{\diff \pperp}{\diff\tau}&=&\frac{e}{mc}\left(\gamma E+i\ppar B\right)
\\
\frac{\diff \gamma}{\diff\tau}&=&\frac{e}{mc}
\textrm{Re}\left(\pperp E^*\right)
\label{circgamma}
\eqe
(where $\tau$ is the proper time ($\diff t=\gamma\diff\tau$) and 
${\diff}/{\diff\tau}={\partial}/{\partial \tau}+ 
c \ppar {\partial }/{\partial r}$
is the convective
derivative),
together with Faraday's law and Amp\`ere's law:
\eqb
\frac{1}{r}\frac{\partial }{\partial r}\left(rE\right)-\frac{i}{c}\frac{\partial B}{\partial t}
&=&0
\\
\frac{1}{r}\frac{\partial }{\partial r}\left(rB\right)+\frac{i}{c}\frac{\partial E}{\partial t}
&=&-i8\pi n e \pperp
\eqe

In the limit of large $r$, approximate solutions to these equations
can be found that describe nonlinear waves, 
using the method of \cite{akhiezerpolovin56}. Introducing the  
phase
\eqb
\phi&=&\omega\left[t-\frac{1}{c}\int_0^r\frac{\diff r'}{\beta\left(r'\right)}\right]
\eqe
one can identify a circularly polarised solution 
valid for superluminal phase speeds ($\beta>1$). In it, the fields are
\eqb
B&=&iE/\beta;\qquad E\,=\,\left|E\right|\textrm{e}^{\pm i\phi}
\eqe
and the two fluids have oppositely directed, non-zero
components of momentum perpendicular to the propagation direction,
whose magnitude is related to the electric field strength:
\eqb
\left|\pperp\right|^2
&=&\frac{e^2\left|E\right|^2}{m^2c^2\omega^2}
\\
&=&\gamma^2-\ppar^2-1
\eqe
where $\left|\pperp\right|$ and $\left|E\right|$ are constants.
This extra degree of freedom 
(compared to the subluminal striped wind) brings with it a restriction on 
the plasma density in the form of a non-linear dispersion
relation:
\eqb
\omega&=&\gamma_{\rm EM}\omega_{\rm p}
\label{dispersion}
\eqe
where $\omega_{\rm p}=\left(8\pi n e^2/m\right)^{1/2}$,
and $\gamma_{\rm EM}=\beta/\sqrt{\beta^2-1}$ is the Lorentz factor that 
corresponds to a velocity of $c/\beta$. Seen from a frame moving 
radially outwards with this speed, the wave is spatially homogeneous
and the magnetic field vanishes. 

\subsection{Perturbation expansion}

Applying a short wavelength approximation to the above equations 
reveals that the slow radial evolution of the wave amplitude is governed by
the conservation of particle number, energy and radial momentum.
\eqb
C_J&=&\hat{r}^2\left<J\right>\,=\,\hat{r}^2 2cn\ppar
\label{Jflux}
\\
C_F&=&\frac{\hat{r}^2\left<F\right>}{mc^2}\,=\,
\hat{r}^2\left(2nc\gamma\ppar+\frac{\left|E\right|^2}{4\pi mc\beta}\right)
\label{Fflux}
\\
C_Q&=&\frac{\hat{r}^2\left<Q\right>}{mc}\,=\,
\hat{r}^2\left[
2nc\ppar^2+\frac{\left|E\right|^2}{8\pi mc}\left(1+\frac{1}{\beta^2}\right)
\right]
\label{Qflux}
\eqe

The dispersion relation (\ref{dispersion}) implies that the proper 
density is proportional to $\gamma_{\rm EM}^{-2}$. Therefore, because of 
the conservation of particle flux, $\ppar\propto \gamma_{\rm EM}^2/\hat{r}^2$.
The constant of proportionality is not important for the analysis of
the wave properties, but it can be found from the spin-down power:
\eqb
\ppar&=&\frac{a_{\rm L}^2\gamma_{\rm EM}^2}{\mu\hat{r}^2}
\label{ppareq}
\eqe
where $a_{\rm L}$ is the strength parameter defined in (\ref{aldefinition}).
As a result of (\ref{ppareq}), this wave propagates
only at a sufficiently large distance from the star. Since $\ppar<\gamma<\mu$, 
the wave propagates only for
\eqb
\hat{r}&>&\frac{a_{\rm L}\gamma_{\rm EM}}{\mu}
\,>\,\frac{a_{\rm L}}{\mu}
\nonumber\\
&=&
4.3\times 10^{6} \mu_4^{-1}L_{38}^{1/2}
\left(4\pi/\Omega_{\rm s}\right)^{1/2}
\eqe
where $\mu_4=\mu/10^4$. 
In the following, the dimensionless radius will be expressed in terms 
of this minimum radius for propagation, using the variable
\eqb
R&=&\hat{r}\mu/a_{\rm L}
\eqe

From (\ref{Jflux}) and (\ref{Fflux}) the $\mu$ parameter is
\eqb
\mu&=&\frac{\left<F\right>}{mc^2 \left<J\right>}
\nonumber\\
&=&\gamma+\frac{\left|E\right|^2}{8\pi \beta n m c^2\ppar}
\\
&=&\gamma+\frac{\gamma_{\rm EM}^2\pperp^2}{\beta \ppar}
\label{mueq}
\eqe
where the dispersion relation (\ref{dispersion}) has been used in the 
last step. 
Similarly, from (\ref{Jflux}) and (\ref{Qflux}) one finds
\eqb
\frac{C_Q}{C_J}&=&
\ppar+\frac{\gamma_{\rm EM}^2\pperp^2}{2\ppar}\left(1+\frac{1}{\beta^2}\right)
\label{emeq2}
\eqe
The radial dependence of the wave 
follows from (\ref{mueq})
and (\ref{emeq2}), together with 
\eqb
\ppar&=&\frac{\gamma_{\rm EM}^2\mu}{R^2}
\label{pparReq}
\\
\gamma^2&=&1+\ppar^2+\pperp^2
\\
\beta&=&\frac{\gamma_{\rm EM}}{\sqrt{\gamma_{\rm EM}^2-1}}
\eqe

\begin{figure}
\includegraphics[bb=50 50 228 175,width=\textwidth]{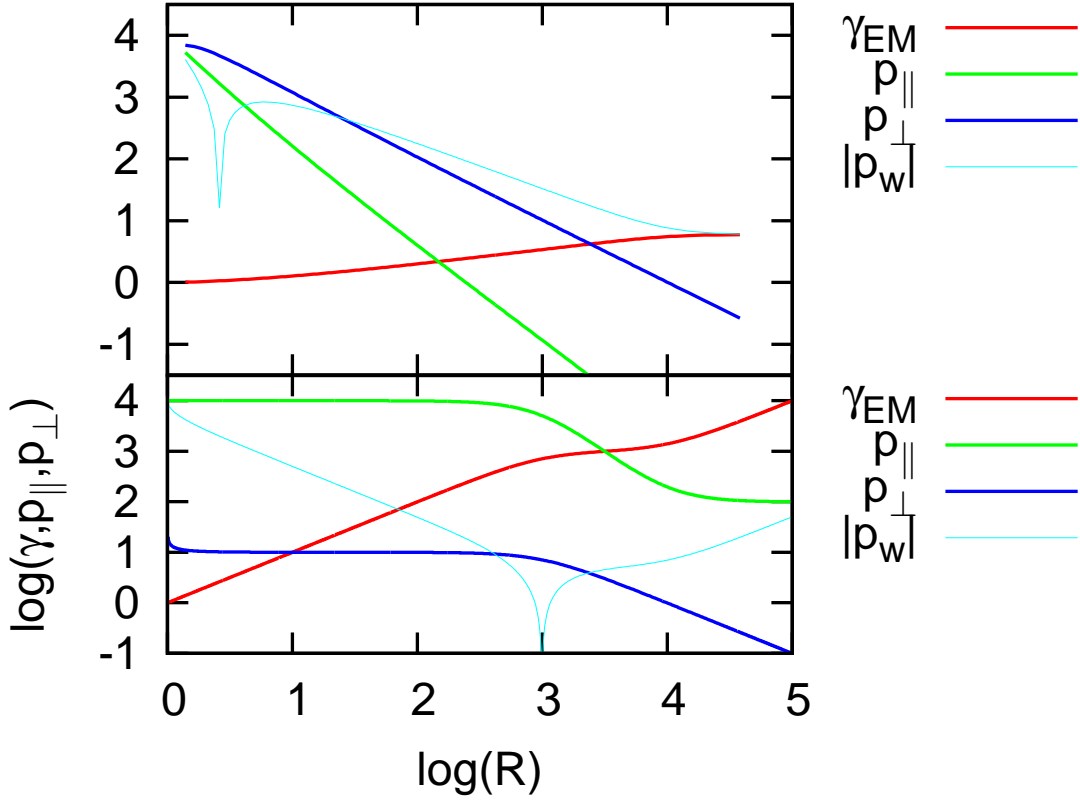}
\caption{\label{emfigure}
The properties of circularly polarised EM waves for $\mu=10^4$, 
$\sigma=10^2$. The quantity $p_{\rm w}$ is the radial momentum of the 
electron and positron fluids seen from the wave frame. It is positive
(wave slower than particles) at $R=1$, but becomes negative at large $R$.}
\end{figure}

Using (\ref{emeq2}) to eliminate $\pperp$ in (\ref{mueq}) enables one to find
a quadratic equation for $\ppar$, so that for a given wave speed
$\gamma_{\rm EM}$, there are at most two distinct physical solutions for 
$\ppar$ and, hence, for the radius. 
In order to relate these solutions 
to the striped wind solutions that carry the same 
particle, energy and momentum fluxes, it suffices to
express $C_Q/C_J$ in terms of the magnetisation parameter
$\sigma$ using (\ref{mhdcqcj})
The radial dependence of these two modes 
for $\mu=10^4$ and $\sigma=10^2$
is shown in Fig.~\ref{emfigure}.

To find the behaviour at large radius, first write
an expression for $\pperp$, using (\ref{emeq2}) and 
(\ref{pparReq}):
\eqb
\pperp^2&=&\frac{2\mu\gamma_{\rm EM}^2}{R^2\left(2\gamma_{\rm EM}^2-1\right)}
\left[\mu\left(1-\frac{\gamma_{\rm EM}^2}{R^2}\right)-
\frac{1+\sigma}{2\mu}\right]
\eqe
As $R\rightarrow\infty$, $\gamma_{\rm EM}$ cannot rise faster than $R$, 
since $\ppar<\mu$. Therefore, for both modes, 
$\pperp\sim R^{-1}$, i.e., the 
wave amplitude decreases as $1/R$. 

\subsection{Free-escape mode}
Writing $\gamma_{\rm EM}=y R$ and expanding in powers of $1/R$
gives
\eqb
y&=&\frac{1}{\sqrt{1+\sigma}}
  \left[1-\left(\frac{1+\sigma}{2\mu}\right)^2\right]^{1/2}
\ \textrm{as}\ R\rightarrow\infty
\eqe
Assuming $\mu/(1+\sigma)\gg1$, this mode has the properties
\eqb
\gamma_{\rm EM}&\rightarrow&R/\sqrt{1+\sigma}
\\
\pperp&\rightarrow&\frac{\mu\sqrt{\sigma}}{R\left(1+\sigma\right)^{1/2}}
\\
\ppar&\rightarrow&\frac{\mu}{1+\sigma}
\eqe
as $R\rightarrow\infty$.
At large distances, its Lorentz factor increases linearly with 
$R$, but the radial momentum of the particles tends to a constant,
so that the density falls of as  $R^{-2}$. The wave essentially 
detaches itself from the particles, and becomes a vacuum wave with a
frequency large compared to the local plasma frequency. 
Most of the 
energy flux at large radius is carried by the wave, in the sense that the 
ratio of the two terms on the RHS of (\ref{mueq}) is 
$\left[
\gamma_{\rm EM}^2\pperp^2/\left(\beta\ppar\right)\right]/\gamma=\sigma\gg1$. Thus, the ratio of electromagnetic to particle
energy flux tends to the same value as in the MHD wave that carries the 
same total particle, energy and momentum fluxes.

At smaller radius, this mode changes character, roughly at the point where 
the wave Lorentz factor equals that of the particles, 
i.e., $R\approx\mu/\sqrt{1+\sigma}$. Inside this point, the wave moves 
outwards more slowly than the particles. Since $\ppar\approx\gamma\approx\mu$,
most
of the energy flux in this inner region is carried by the 
particles. The wave amplitude $\pperp$ is nearly constant.

\subsection{Confined mode}
The second mode is very different. At large radius,
$\gamma_{\rm EM}$ tends to a constant, 
$\ppar\rightarrow0$ and $\gamma\rightarrow1$.
In this case, combining (\ref{mueq}) and (\ref{emeq2}) 
for large $R$ leads to 
\eqb
\mu-\frac{1+\sigma}{2\mu}&=&\beta(\mu-1)\frac{1}{2}\left(1+\frac{1}{\beta^2}\right)
\eqe
This equation has one superluminal ($\beta>1$) root provided
$\mu>(1+\sigma)/2$, as is expected under pulsar conditions.
For large $\mu$ it corresponds to 
\eqb
\gamma_{\rm EM}&\rightarrow&\left(\mu/8\right)^{1/4}
\\
\pperp&\rightarrow&\mu/R
\\
\ppar&\rightarrow&\mu^{3/2}/\left(R^2\sqrt{8}\right)
\eqe

Close to $R=1$, this wave moves more slowly than the particles, but 
quickly overtakes them. However, it does not detach itself. 
The radial momentum of the particles decreases as $R^{-2}$, so that
the proper density remains constant. 
The ratio of the wave frequency to the local plasma frequency 
approaches the value $\left(\mu/8\right)^{1/4}$.  
Most of the energy flux is carried
by the wave --- for this mode at large radius
the ratio of the terms in (\ref{mueq}) equals $\mu$, substantially greater 
than the ratio in the equivalent MHD wave.

\section{Summary}
An analysis is presented of some simple  
nonlinear waves that 
propagate in the inner and outer parts of a pulsar wind.
These are described by the equations of relativistic 
MHD in the inner part, and by a model with two relativistic
cold fluids (electron and positron) in the outer part.  
The inner solution tends to convert energy flux carried by the fields
into particle energy flux as it accelerates outwards. Two electromagnetic
outer solutions were found, both of which do the opposite, converting
particle energy flux to wave energy flux. One of the outer solutions
tends to a vacuum wave at large radius 
and has the same ratio of 
particle energy flux to flux carried by the fields as does the 
corresponding MHD wave. The other remains a plasma wave at large
radius, but carries an even larger fraction of its energy flux in the 
fields.

The most important property of the outer wave modes is likely to 
be their damping rates. Previous analyses of the linearly polarised
plane wave case have found strong damping by radiation reaction for
pulsar parameters \cite{asseokennelpellat78}, and, for general parameters,
strong growth of the Weibel instability driven by the 
counter-streaming fluids \cite{asseollobetschmidt80}. 
These and similar effects are likely to 
determine the structure of the \lq\lq termination shock\rq\rq\ at the 
interface between the pulsar wind and the nebula, but have yet to be analysed 
in a global model.
\ack
I thank A.~Bell, I.~Mochol, J.~P\'etri and B.~Reville for helpful discussions.

\end{document}